\documentclass[12pt,a4paper,english]{article}
\usepackage[T1]{fontenc}
\usepackage[latin9]{inputenc}
\usepackage{amssymb}

\makeatletter

\usepackage{a4wide}
\makeatletter

\@addtoreset{equation}{section}

\@addtoreset{figure}{section}

\@addtoreset{table}{section}
\makeatother

\usepackage{babel}
\makeatother

\begin{document}

\title{\begin{flushright}{\normalsize ITP-Budapest Report No. 637}\end{flushright}\vspace{1cm}Form
factors of boundary exponential operators in the sinh-Gordon model}

\author{G. Takács%
\thanks{E-mail: takacs@elte.hu%
}\\
{\small Theoretical Physics Research Group of the Hungarian Academy
of Sciences}\\
{\small H-1117 Budapest, Pázmány Péter sétány 1/A}}

\date{7th January, 2008}

\maketitle
\begin{abstract}
Using the recently introduced boundary form factor bootstrap equations,
the form factors of boundary exponential operators in the sinh-Gordon
model are constructed. The ultraviolet scaling dimension and the normalization
of these operators are checked against previously known results. The
construction presented in this paper can be applied to determine form
factors of relevant primary boundary operators in general integrable
boundary quantum field theories.
\end{abstract}

\section{Introduction}

The investigation of integrable boundary quantum field theories started
with the seminal work of Ghoshal and Zamolodchikov \cite{GZ}, who
set up the boundary R-matrix bootstrap, which makes possible the determination
of the reflection matrices and provides complete description of the
theory on the mass shell.

For the calculation of correlation functions, matrix elements of local
operators between asymptotic states have to be computed. In a boundary
quantum field theory there are two types of operators, the bulk and
the boundary operators, where their names indicate their localization
point. The boundary bootstrap program, namely the boundary form factor
program for calculating the matrix elements of local boundary operators
between asymptotic states was initiated in \cite{bffprogram}. The
validity of form factor solutions was checked in the case of the boundary
scaling Lee-Yang model calculating the two-point function using a
spectral sum and comparing it to the prediction of conformal perturbation
theory. In \cite{bffcount} the spectrum of independent form factor
solutions was compared to the boundary operator content of the ultraviolet
boundary conformal field theory and a complete agreement was found.
Further solutions of the boundary form factor axioms were constructed
and their structure was analyzed for the sinh-Gordon theory at the
self-dual point in \cite{ca1}, and for the $A_{2}$ affine Toda field
theory in \cite{ca2}. In the recent paper \cite{bfftcsa} the validity
of the form factor solution conjectured for the unique nontrivial
boundary primary field in scaling Lee-Yang model was tested against
truncated conformal space and a spectacular agreement was found. 

It is clear from the discussion in \cite{bffcount} that the most
interesting open problem of the boundary form factor bootstrap is
the identification of the operator corresponding to a given solution.
For example, in sinh-Gordon theory there are infinitely many form
factor solutions with minimal growth at large rapidities, which can
be attributed to the presence of exponential boundary fields. The
task undertaken in this paper is to map the space of such minimal
solutions, and make their correspondence with exponential fields more
precise. It is shown that specific solutions can be selected inside
this infinite family, such that their scaling dimension agrees with
the prediction of conformal field theory, while their normalization
matches the available results on the vacuum expectation value of exponential
operators. The construction of these solutions can be generalized
to determine form factors of relevant boundary primary fields in any
model where the form factors of relevant operators in the bulk theory
are known.

The outline of the paper is the following. In section 2 the necessary
information about boundary sinh-Gordon theory is presented. Section
3 describes the construction of the form factors solutions which are
conjectured to correspond to boundary exponential fields. In section
4 their ultraviolet dimension and normalization is evaluated as a
series expansion in the bulk parameter of sinh-Gordon theory, and
is shown to be consistent with known results. Section 5 is reserved
for the conclusions.

\section{Boundary sinh-Gordon theory}

The sinh-Gordon theory in the bulk is defined by the Lagrangian density
\[
\mathcal{L}=\frac{1}{2}(\partial_{\mu}\Phi)^{2}-\frac{m^{2}}{b^{2}}(\cosh b\Phi-1)\]
It can be considered as the analytic continuation of the sine-Gordon
model for imaginary coupling $\beta=ib$. The S-matrix of the model
is\[
S(\theta)=-\left(1+\frac{B}{2}\right)_{\theta}\left(-\frac{B}{2}\right)_{\theta}=\left[-\frac{B}{2}\right]_{\theta}\qquad;\quad B=\frac{2b^{2}}{8\pi+b^{2}}\]
where \[
(x)_{\theta}=\frac{\sinh\frac{1}{2}\left(\theta+i\pi x\right)}{\sinh\frac{1}{2}\left(\theta-i\pi x\right)}\quad,\quad[x]_{\theta}=-(x)_{\theta}(1-x)_{\theta}=\frac{\sinh\theta+i\sin\pi x}{\sinh\theta-i\sin\pi x}\]
 The minimal bulk two-particle form factor belonging to this S-matrix
is \cite{FMS}\begin{equation}
f(\theta)=\mathcal{N}\exp\left[8\int_{0}^{\infty}\frac{dx}{x}\sin^{2}\left(\frac{x(i\pi-\theta)}{2\pi}\right)\frac{\sinh\frac{xB}{4}\sinh(1-\frac{B}{2})\frac{x}{2}\sinh\frac{x}{2}}{\sinh^{2}x}\right]\label{eq:fmin}\end{equation}
where \begin{equation}
\mathcal{N}=\exp\left[-4\int_{0}^{\infty}\frac{dx}{x}\frac{\sinh\frac{xB}{4}\sinh(1-\frac{B}{2})\frac{x}{2}\sinh\frac{x}{2}}{\sinh^{2}x}\right]\label{eq:NN}\end{equation}
It satisfies $f(\theta,B)\rightarrow1$ as $\theta\rightarrow\infty$,
and approaches its asymptotic value exponentially fast. 

Sinh-Gordon theory can be restricted to the negative half-line with
the following action\begin{eqnarray}
\mathcal{A} & = & \int_{-\infty}^{\infty}dt\int_{-\infty}^{0}dx\left[\frac{1}{2}(\partial_{\mu}\Phi)^{2}-\frac{m^{2}}{b^{2}}(\cosh b\Phi-1)\right]\label{eq:boundaryaction}\\
 &  & +\int_{-\infty}^{\infty}dtM_{0}\left(\cosh\left(\frac{b}{2}(\Phi(0,t)-\Phi_{0})\right)-1\right)\nonumber \end{eqnarray}
which maintains integrability \cite{GZ}. The corresponding reflection
factor depends on two continuous parameters and can be written as
\cite{shGEF}\begin{equation}
R(\theta)=\left(\frac{1}{2}\right)_{\theta}\left(\frac{1}{2}+\frac{B}{4}\right)_{\theta}\left(1-\frac{B}{4}\right)_{\theta}\left[\frac{E-1}{2}\right]_{\theta}\left[\frac{F-1}{2}\right]_{\theta}\label{eq:shgrefl}\end{equation}
It can be obtained as the analytic continuation of the first breather
reflection factor in boundary sine-Gordon model which was calculated
by Ghoshal in \cite{ghoshal}. The relation of the bootstrap parameters
$E$ and $F$ to the parameters of the Lagrangian is known both from
a semi-classical calculation \cite{shGEF,corrigan} and also in an
exact form in the perturbed boundary conformal field theory framework
\cite{sinG_uvir}.

\section{Boundary form factors in sinh-Gordon theory }

\subsection{The boundary form factor axioms}

The axioms satisfied by the form factors of a local boundary operator
were derived in \cite{bffprogram} and are listed here without much
further explanation. Let us assume that the spectrum contains a single
scalar particle of mass $m$, which has a two-particle $S$ matrix
$S(\theta)$ (using the standard rapidity parametrization) and a one-particle
reflection factor $R(\theta)$ off the boundary, satisfying the boundary
reflection factor bootstrap conditions of Ghoshal and Zamolodchikov
\cite{GZ}. For a local operator $\mathcal{O}(t)$ localized at the
boundary (located at $x=0$, and parametrized by the time coordinate
$t$) the form factors are defined as\begin{eqnarray*}
\,_{out}\langle\theta_{1}^{'},\theta_{2}^{'},\dots,\theta_{m}^{'}\vert\mathcal{O}(t)\vert\theta_{1},\theta_{2},\dots,\theta_{n}\rangle_{in} & =\\
 &  & \hspace{-2cm}F_{mn}^{\mathcal{O}}(\theta_{1}^{'},\theta_{2}^{'},\dots,\theta_{m}^{'};\theta_{1},\theta_{2},\dots,\theta_{n})e^{-imt(\sum\cosh\theta_{i}-\sum\cosh\theta_{j}^{'})}\end{eqnarray*}
for $\theta_{1}>\theta_{2}>\dots>\theta_{n}>0$ and $\theta_{1}^{'}<\theta_{2}^{'}<\dots<\theta_{m}^{'}<0$,
using the asymptotic $in/out$ state formalism introduced in \cite{BBT}.
They can be extended analytically to other values of rapidities. With
the help of the crossing relations derived in \cite{bffprogram} all
form factors can be expressed in terms of the elementary form factors\[
\,_{out}\langle0\vert\mathcal{O}(0)\vert\theta_{1},\theta_{2},\dots,\theta_{n}\rangle_{in}=F_{n}^{\mathcal{O}}(\theta_{1},\theta_{2},\dots,\theta_{n})\]
which can be shown to satisfy the following axioms: 

I. Permutation:

\begin{center}
\[
F_{n}^{\mathcal{O}}(\theta_{1},\dots,\theta_{i},\theta_{i+1},\dots,\theta_{n})=S(\theta_{i}-\theta_{i+1})F_{n}^{\mathcal{O}}(\theta_{1},\dots,\theta_{i+1},\theta_{i},\dots,\theta_{n})\]

\par\end{center}

II. Reflection:\[
F_{n}^{\mathcal{O}}(\theta_{1},\dots,\theta_{n-1},\theta_{n})=R(\theta_{n})F_{n}^{\mathcal{O}}(\theta_{1},\dots,\theta_{n-1},-\theta_{n})\]

III. Crossing reflection: \[
F_{n}^{\mathcal{O}}(\theta_{1},\theta_{2},\dots,\theta_{n})=R(i\pi-\theta_{1})F_{n}^{\mathcal{O}}(2i\pi-\theta_{1},\theta_{2},\dots,\theta_{n})\]

IV. Kinematical singularity\[
-i\mathop{\textrm{Res}}_{\theta=\theta^{'}}F_{n+2}^{\mathcal{O}}(\theta+i\pi,\theta^{'},\theta_{1},\dots,\theta_{n})=\left(1-\prod_{i=1}^{n}S(\theta-\theta_{i})S(\theta+\theta_{i})\right)F_{n}^{\mathcal{O}}(\theta_{1},\dots,\theta_{n})\]

V. Boundary kinematical singularity

\[
-i\mathop{\textrm{Res}}_{\theta=0}F_{n+1}^{\mathcal{O}}(\theta+\frac{i\pi}{2},\theta_{1},\dots,\theta_{n})=\frac{g}{2}\Bigl(1-\prod_{i=1}^{n}S\bigl(\frac{i\pi}{2}-\theta_{i}\bigr)\Bigr)F_{n}^{\mathcal{O}}(\theta_{1},\dots,\theta_{n})\]
where $g$ is the one-particle coupling to the boundary\begin{equation}
R(\theta)\sim\frac{ig^{2}}{2\theta-i\pi}\quad,\quad\theta\sim i\frac{\pi}{2}\label{eq:gdef}\end{equation}

There are also axioms for singularities corresponding to bound states
(bulk and boundary), but they are not needed in the sequel.

The general solution of the above axioms can be written as \cite{bffprogram}\begin{equation}
F_{n}(\theta_{1},\theta_{2},\dots,\theta_{n})=H_{n}\frac{Q_{n}(y_{1},y_{2}\dots,y_{n})}{\prod_{i}y_{i}\,\prod\limits _{i<j}(y_{i}+y_{j})}\prod_{i=1}^{n}r(\theta_{i})\prod_{i<j}f(\theta_{i}-\theta_{j})f(\theta_{i}+\theta_{j})\label{eq:GenAnsatz}\end{equation}
where the $Q_{n}$ are symmetric polynomials of its variables, and
the minimal one-particle form factor is given by\begin{equation}
r(\theta)=\frac{i\sinh\theta}{(\sinh\theta-i\sin\gamma)(\sinh\theta-i\sin\gamma')}u(\theta,B)\quad,\quad\gamma=\frac{\pi}{2}(E-1)\quad\gamma'=\frac{\pi}{2}(F-1)\label{eq:rmin}\end{equation}
with asymptotics $r(\theta)\sim1$ when $\theta\rightarrow\infty$,
where%
\footnote{Notice that the normalization used here differs from that in the earlier
papers \cite{bffprogram,bffcount}. %
} \begin{eqnarray*}
u(\theta) & = & \exp\Bigg\{\int_{0}^{\infty}\frac{dt}{t}\left[\frac{1}{\sinh\frac{t}{2}}-2\cosh\frac{t}{2}\cos\left[\left(\frac{i\pi}{2}-\theta\right)\frac{t}{\pi}\right]\right]\times\\
 &  & \frac{\sinh\frac{xB}{4}+\sinh\left(1-\frac{B}{2}\right)\frac{x}{2}+\sinh\frac{x}{2}}{\sinh^{2}t}\Bigg\}\end{eqnarray*}
and \begin{equation}
H_{n}=\left(\frac{4\sin\pi B/2}{f(i\pi)}\right)^{n/2}\label{eq:Hn}\end{equation}
is a convenient normalization factor. Using the results of \cite{bffprogram,bffcount}
it is easy to derive the recursion relations satisfied by the polynomials
$Q_{n}$: \begin{eqnarray}
\mathcal{K}: &  & Q_{2}(-y,y)=0\nonumber \\
 &  & Q_{n+2}(-y,y,y_{1},\dots,y_{n})=\nonumber \\
 &  & (y^{2}-4\cos^{2}\gamma)(y^{2}-4\cos^{2}\gamma')P_{n}(y|y_{1},\dots,y_{n})\, Q_{n}(y_{1},\dots,y_{n})\quad\mbox{for }n>0\nonumber \\
\mathcal{B}: &  & Q_{1}(0)=0\nonumber \\
 &  & Q_{n+1}(0,y_{1},\dots,y_{n})=\nonumber \\
 &  & 4\cos\gamma\cos\gamma'B_{n}(y_{1},\dots,y_{n})\, Q_{n}(y_{1},\dots,y_{n})\quad\mbox{for }n>0\label{eq:sinhgrecursions}\end{eqnarray}
where\begin{equation}
B_{n}(y_{1},\dots,y_{n})=\frac{1}{4\sin\frac{\pi B}{2}}\left(\prod_{i=1}^{n}\left(y_{i}-2\sin\frac{\pi B}{2}\right)-\prod_{i=1}^{n}\left(y_{i}+2\sin\frac{\pi B}{2}\right)\right)\label{eq:Bequ}\end{equation}
and \begin{equation}
P_{n}(y|y_{1},\dots y_{n})=\frac{1}{2(y_{+}-y_{-})}\left[\prod_{i=1}^{n}(y_{i}-y_{-})(y_{i}+y_{+})-\prod_{i=1}^{n}(y_{i}+y_{-})(y_{i}-y_{+})\right]\label{eq:Pequ}\end{equation}
with the notations\begin{eqnarray}
y_{+} & = & \omega z+\omega^{-1}z^{-1}\label{eq:ypmeq}\\
y_{-} & = & \omega^{-1}z+\omega z^{-1}\qquad,\qquad\omega=e^{i\pi\frac{B}{2}}\nonumber \end{eqnarray}
with the auxiliary variable $z$ defined as a solution of $y=z+z^{-1}$
(i.e. writing $y=2\cosh\theta$ one has $z=\mathrm{e}^{\theta}$). 

The two-point functions can be computed from a spectral representation:\begin{eqnarray}
\rho_{\mathcal{AB}}(mt) & = & \langle0\vert\mathcal{A}(t)\mathcal{B}(0)\vert0\rangle\label{eq:2pt}\\
 & = & \sum_{n=0}^{\infty}\frac{1}{(2\pi)^{n}}\int_{\theta_{1}>\dots>\theta_{n}>0}d\theta_{1}\dots d\theta_{n}\, f_{n}\left(\theta_{1},\dots,\theta_{n}\right)\exp\left(-imt{\displaystyle \sum_{i=1}^{n}}\cosh\theta_{i}\right)\nonumber \end{eqnarray}
where time translation invariance was used and \[
f_{n}\left(\theta_{1},\dots,\theta_{n}\right)=F_{n}^{\mathcal{A}}(\theta_{1},\dots,\theta_{n})^{\dagger}F_{n}^{\mathcal{B}}(\theta_{1},\dots,\theta_{n})\]
The $f_{n}$ are symmetric (and also even) in all their variables,
therefore the spectral expansion (\ref{eq:2pt}) can be written in
the following form for the Euclidean two-point function\begin{eqnarray*}
\rho_{\mathcal{AB}}(m\tau) & = & \sum_{n=0}^{\infty}\frac{1}{n!}\frac{1}{(2\pi)^{n}}\int_{0}^{\infty}d\theta_{1}\int_{0}^{\infty}d\theta_{2}\dots\int_{0}^{\infty}d\theta_{n}e^{-m\tau\sum_{i}\cosh\theta_{i}}f_{n}\left(\theta_{1},\dots,\theta_{n}\right)\end{eqnarray*}
The operators of interest can be classified according to their scaling
dimensions, which means that the two-point function must have a power-like
short-distance singularity\begin{equation}
\rho_{\mathcal{AB}}(m\tau)=\frac{1}{\tau^{2\Delta_{\mathcal{AB}}}}+\dots\label{eq:shortdistance}\end{equation}
where $\Delta_{\mathcal{AB}}$ is an exponent determined by the ultraviolet
scaling weights of the local fields.

\subsection{The family of minimal solutions and the cumulant expansion}

In the earlier paper \cite{bffcount} it was shown that there exist
an infinite family of solutions for which the polynomials $Q_{n}$
have the minimum possible degree\[
\mathrm{deg}\, Q_{n}=\frac{n(n+1)}{2}\]
These can be thought to correspond to the exponential operators\[
\mathrm{e}^{\alpha\Phi(t,x=0)}\]
of which only a countably infinite number is independent since they
can all be expanded  in terms of powers of the field:\[
\mathrm{e}^{\alpha\Phi(t,0)}=\sum_{k=1}^{\infty}\frac{\alpha^{k}}{k\,!}\Phi^{k}(t,0)\]
For any minimal solution the corresponding form factor has a finite
limit when all rapidities are taken to infinity simultaneously\begin{equation}
F_{n}(\theta_{1}+\lambda,\theta_{2}+\lambda,\dots,\theta_{n}+\lambda)\:\rightarrow\:\tilde{F}_{n}(\theta_{1},\theta_{2},\dots,\theta_{n})+O(\mathrm{e}^{-\lambda})\label{eq:limitingff}\end{equation}
which means that every multi-particle contribution in the spectral
expansion (\ref{eq:2pt}) individually behaves as\[
\tau^{-2\delta}\]
with the naive scaling dimension $\delta$ equal to $0$ \cite{bffprogram,bffcount}. 

However, the true scaling dimension in general turns out to be different
from $0$ due to logarithmic corrections in the individual multi-particle
contributions, which may sum up to give an anomalous dimension. It
can be computed using the cumulant expansion of the logarithm of the
two-point function \cite{smirnovcumulant} (for a very nice discussion
see also \cite{babujiankarowski}). Consider the conformal operator
product expansion\[
\mathcal{A}(\tau)\mathcal{B}(0)\sim\sum_{h_{i}}\frac{C_{\mathcal{AB}}^{i}}{\tau^{h_{\mathcal{A}}+h_{\mathcal{B}}-h_{i}}}\mathcal{O}_{i}(0)\]
where $h_{\mathcal{A}}$ and $h_{\mathcal{B}}$ are the ultraviolet
weights of the fields $\mathcal{A}$ and $\mathcal{B}$, while the
$h_{i}$ are the weight of the $\mathcal{O}_{i}$. It is obvious that\[
2\Delta_{\mathcal{AB}}=h_{\mathcal{A}}+h_{\mathcal{B}}-h_{\mathrm{min}}\]
where $h_{\mathrm{min}}$ is the minimum of the weights $h_{i}$ of
the operators $\mathcal{O}_{i}$ appearing in the expansion. Let us
suppose that the limiting function (\ref{eq:limitingff}) satisfies
an asymptotic factorization property of the form \begin{equation}
\tilde{F}_{n}(\theta_{1},\dots,\theta_{k},\theta_{k+1}+\lambda,\dots,\theta_{n}+\lambda)=\tilde{F}_{k}(\theta_{1},\dots,\theta_{k})\tilde{F}_{n-k}(\theta_{k+1},\dots,\theta_{n})+O(\mathrm{e}^{-\lambda})\label{eq:ffcluster}\end{equation}
both for the form factors of $\mathcal{A}$ and $\mathcal{B}$. The
leading $n=0$ term in the spectral expansions is a constant given
by the vacuum expectation value of the field, which is assumed to
be $1$. In the case of bulk form factors (\ref{eq:ffcluster}) with
the particular normalization given above entails that the operator
has a unit vacuum expectation value \cite{delta-th}. For boundary
form factors this can be carried over as a reasonable assumption which
was used previously to normalize the minimal form factor of the boundary
Lee-Yang model in \cite{bffprogram}. This assumption is reasonable
from the consistency of the arguments in this paper and is also checked
explicitly in subsection 4.2.

Under the above assumptions, the logarithm of the correlation function
can be written as\begin{equation}
\log\rho_{\mathcal{AB}}(m\tau)=\sum_{n=1}^{\infty}\frac{1}{n!}\frac{1}{(2\pi)^{n}}\int_{0}^{\infty}d\theta_{1}\int_{0}^{\infty}d\theta_{2}\dots\int_{0}^{\infty}d\theta_{n}\mathrm{e}^{-m\tau\sum_{i}\cosh\theta_{i}}c_{n}\left(\theta_{1},\dots,\theta_{n}\right)\label{eq:logrho}\end{equation}
where the $c_{n}$ are the cumulants of the functions $f_{n}$ defined
recursively by\begin{eqnarray*}
 &  & f_{1}(\theta_{1})=c_{1}(\theta_{1})\quad,\quad f_{2}(\theta_{1},\theta_{2})=c_{2}(\theta_{1},\theta_{2})+c_{1}(\theta_{1})c_{1}(\theta_{2})\\
 &  & f_{3}(\theta_{1},\theta_{2},\theta_{3})=c_{3}(\theta_{1},\theta_{2},\theta_{3})+c_{1}(\theta_{1})c_{2}(\theta_{2},\theta_{3})+c_{1}(\theta_{2})c_{2}(\theta_{1},\theta_{3})+c_{1}(\theta_{3})c_{2}(\theta_{1},\theta_{2})\\
 &  & +c_{1}(\theta_{1})c_{1}(\theta_{2})c_{1}(\theta_{3})\\
 &  & \dots\end{eqnarray*}
Defining \begin{equation}
\tilde{c}_{n}(\theta_{1},\dots,\theta_{n})=\lim_{\lambda\rightarrow\infty}c_{n}(\theta_{1}+\lambda,\dots,\theta_{n}+\lambda)\label{eq:hbarlimit}\end{equation}
it is easy to see that the functions $\tilde{c}_{n}$ depend only
on the differences of the rapidities. From (\ref{eq:ffcluster}) it
is easy to obtain the following property of the asymptotic cumulants\begin{equation}
\tilde{c}_{n}(\theta_{1},\dots,\theta_{k},\theta_{k+1}+\lambda,\dots,\theta_{n}+\lambda)\sim O(\mathrm{e}^{-\lambda})\qquad k=1,\dots,n-1\label{eq:cumulantdecay}\end{equation}
and also note that \begin{equation}
c_{n}(\theta_{1}+\lambda,\dots,\theta_{n}+\lambda)=\tilde{c}_{n}(\theta_{1},\dots,\theta_{n})+O(\mathrm{e}^{-\lambda})\label{eq:cumulantapproach}\end{equation}
It can then be shown that \begin{equation}
2\Delta_{\mathcal{A}B}=\sum_{n=1}^{\infty}\frac{1}{n!}\frac{1}{(2\pi)^{n}}\int_{-\infty}^{\infty}d\theta_{2}\dots\int_{-\infty}^{\infty}d\theta_{n}\tilde{c}_{n}\left(0,\theta_{2},\dots,\theta_{n}\right)\label{eq:cumulantexpansion}\end{equation}
The derivation of this result is a bit more involved than in the bulk
case where the translational invariance of the form factor in rapidity
space can be used. Let us examine the $n=1$ contribution in more
detail. The formulae encountered will also be useful in subsection
4.2. Consider the integral\[
\int_{0}^{\infty}\frac{d\theta}{2\pi}c_{1}(\theta)\mathrm{e}^{-m\tau\cosh\theta}\]
Due to (\ref{eq:cumulantapproach}), $c_{1}(\theta)$ approaches its
limiting value exponentially fast\[
c_{1}(\theta)=\tilde{c}_{1}+O(\mathrm{e}^{-\alpha\theta})\]
The exponential factor has the property\[
\mathrm{e}^{-m\tau\cosh\theta}\sim1\qquad\theta\ll\log\frac{2}{m\tau}\]
and therefore in the limit of small $\tau$ \begin{equation}
\int_{0}^{\infty}\frac{d\theta}{2\pi}c_{1}(\theta)\mathrm{e}^{-m\tau\cosh\theta}\sim\int_{0}^{\infty}\frac{d\theta}{2\pi}\left(c_{1}(\theta)-\tilde{c}_{1}\right)+\int_{0}^{\infty}\frac{d\theta}{2\pi}\tilde{c}_{1}\mathrm{e}^{-m\tau\cosh\theta}\label{eq:1stcum}\end{equation}
which can be evaluated as\[
\mbox{constant}+\frac{\tilde{c}_{1}}{2\pi}K_{0}(m\tau)\]
Using the asymptotics\begin{equation}
K_{0}(m\tau)\sim-\log m\tau+\log2-\gamma_{E}\label{eq:besselas}\end{equation}
(where $\gamma_{E}=0.577215\dots$ is Euler's constant) the short-distance
exponent is given in this approximation as\begin{equation}
2\Delta_{\mathcal{AB}}=\frac{\tilde{c}_{1}}{2\pi}+\dots\label{eq:1stcumulant}\end{equation}
Turning to $n>1$, the expansion (\ref{eq:logrho}) can be rewritten
as \[
\log\rho_{\mathcal{AB}}(m\tau)=\sum_{n=1}^{\infty}\frac{1}{n!}\frac{1}{(2\pi)^{n}}\frac{1}{2^{n}}\int_{-\infty}^{\infty}d\theta_{1}\int_{-\infty}^{\infty}d\theta_{2}\dots\int_{-\infty}^{\infty}d\theta_{n}\mathrm{e}^{-m\tau\sum_{i}\cosh\theta_{i}}c_{n}\left(\theta_{1},\dots,\theta_{n}\right)\]
using that the functions $f_{n}$ and consequently also $c_{n}$ are
even in all their arguments. For $m\tau\ll1$, the support of the
integrand is concentrated inside the hypercube \[
|\theta_{i}|\lesssim\log(2/m\tau)\]
and in the interior (except for a transient shell-like region whose
thickness is $O(1)$ independent of $\tau$) \[
\mathrm{e}^{-m\tau\sum_{i}\cosh\theta_{i}}\sim1\]
Using the properties (\ref{eq:cumulantdecay}) and (\ref{eq:cumulantapproach})
$c_{n}$ can be replaced by $\tilde{c}_{n}$ up to $O(1)$ terms.
One can now repeat the same procedure in the $\theta_{1}$ integral
as for $n=1$, following the line of the derivation used in the bulk
case \cite{babujiankarowski}. However, some additional care must
be taken since the functions $c_{n}$ are even, and therefore there
are $2^{n-1}$ independent asymptotic directions in which the integrand
is non-vanishing: \[
c_{n}\left(\theta_{1}+\lambda,\theta_{2}+\sigma_{2}\lambda,\dots,\theta_{n}+\sigma_{n}\lambda\right)\rightarrow\tilde{c}_{n}\left(\theta_{1},\sigma_{2}\theta_{2},\dots,\sigma_{n}\theta_{n}\right)\]
depending on the choices of the signs $\sigma_{i}=\pm1$ (the sign
of $\theta_{1}$ is fixed as only the relative signs matter). After
a redefinition of integration variables $\theta_{i}\rightarrow\sigma_{i}\theta_{i}$
this results in $2^{n-1}$ identical contribution which cancel out
$n-1$ factors of $1/2$. The remaining factor $1/2$ can be used
to map back the $\theta_{1}$ integration to the half-line where (\ref{eq:besselas})
can be used.

In the bulk the factor $1/2^{n}$ are absent, and translational invariance
of the form factor of a spinless operator implies that $c_{n}\equiv\tilde{c}_{n}$.
Due to (\ref{eq:cumulantdecay}) only the asymptotic region with all
signs positive contributes, so there is no factor $2^{n-1}$ either.
Therefore the bulk singularity exponent is twice the one in the boundary
case (provided that the asymptotic cumulants are taken to be identical).
However, if the singularity of the bulk two-point function is taken
to be of the form\[
\frac{1}{\tau^{4\Delta_{\mathcal{AB}}}}\]
then this factor cancels out in the final formula for $\Delta_{\mathcal{AB}}$
and so the bulk cumulant expansion is identical to (\ref{eq:cumulantexpansion}).
The additional factor of $2$ in the exponent is natural since in
the bulk the conformal weight receives two identical contributions
from the left and the right movers.

\subsection{The limiting case: bulk form factor solutions in the sinh-Gordon
model}

Now the task is to resolve the ambiguity in the minimal solution.
From \cite{bffcount} it is known to originate from the kernel of
the recursion relations, which is a homogeneous polynomial of order
$n(n+1)/2$ at $n$-particle level. Let us introduce\[
\tilde{Q}_{n}(y_{1},\dots,y_{n})=\lim_{\Lambda\rightarrow\infty}\Lambda^{-n(n+1)/2}Q_{n}(\Lambda y_{1},\dots,\Lambda y_{n})\]
which gives the terms in $Q_{n}$ which are exactly of degree $n(n+1)/2$
(all other terms have lower degree). 

Taking now a solution of the form (\ref{eq:GenAnsatz}), the limiting
procedure (\ref{eq:limitingff}) gives\begin{eqnarray*}
\tilde{F}_{n}(\theta_{1},\theta_{2},\dots,\theta_{n}) & = & \lim_{\lambda\rightarrow\infty}F_{n}(\theta_{1}+\lambda,\theta_{2}+\lambda,\dots,\theta_{n}+\lambda)\\
 & = & H_{n}\frac{\tilde{Q}_{n}(x_{1},x_{2}\dots,x_{n})}{\prod_{i}x_{i}\,\prod\limits _{i<j}(x_{i}+x_{j})}\prod_{i<j}f(\theta_{i}-\theta_{j})\end{eqnarray*}
where $x_{i}=\mathrm{e}^{\theta_{i}}$. It was observed in \cite{bffcount}
that as a consequence of the boundary form factor axioms listed in
subsection 3.1, $\tilde{F}$ is always a solution of the bulk form
factor axioms built upon the same bulk $S$ matrix. For the bulk sinh-Gordon
theory, a complete set of minimal solutions was found by Koubek and
Mussardo in \cite{koubek_mussardo}. Let us define the elementary
symmetric polynomials by\begin{eqnarray*}
\prod_{i=1}^{n}(x+x_{i}) & = & \sum_{l=1}^{n}x^{n-l}\sigma_{l}^{(n)}(x_{1},\dots,x_{n})\\
\sigma_{l}^{(n)}\equiv0 &  & \mbox{if }\: l<0\mbox{ or }l>n\end{eqnarray*}
The upper index will be omitted in the sequel, as the number of variables
will always be clear from the context. Let us also denote\[
[n]=\frac{\omega^{n}-\omega^{-n}}{\omega-\omega^{-1}}=\frac{\sin\frac{n\pi B}{2}}{\sin\frac{\pi B}{2}}\]
Koubek and Mussardo state that the following family of form factor
solutions parametrized by the real number $k$\[
\tilde{F}_{n}^{(k)}(\theta_{1},\theta_{2},\dots,\theta_{n})=H_{n}\frac{P_{n}^{(k)}(x_{1},x_{2}\dots,x_{n})}{\prod\limits _{i<j}(x_{i}+x_{j})}\prod_{i<j}f(\theta_{i}-\theta_{j})\]
 where the polynomials $P_{n}^{(k)}$ are given by\begin{eqnarray*}
P_{1}^{(k)} & = & [k]\\
P_{n}^{(k)} & = & [k]\det M^{(n)}(k)\qquad n>1\\
 &  & M_{ij}^{(n)}(k)=[i-j+k]\sigma_{2i-j}(x_{1},x_{2}\dots,x_{n})\quad i,j=1,\dots,n-1\end{eqnarray*}
correspond to the bulk exponential operators\[
\mathrm{e}^{kg\Phi}\]
normalized so that their vacuum expectation value is unity:\[
\left\langle \mathrm{e}^{kg\Phi}\right\rangle =1\]
They support the identification as follows. The above solutions satisfy
the asymptotic factorization property (\ref{eq:ffcluster}), derived
in \cite{delta-th} for form factors of relevant primary fields. In
addition, their conformal weights to lowest order in $g^{2}$ (computed
from the two-point function) are given by\[
\Delta_{k}=\bar{\Delta}_{k}=-\frac{k^{2}g^{2}}{8\pi}\]
which equal the conformal weights of the exponential operator $\mathrm{e}^{kg\Phi}$
computed in the free boson conformal field theory. In section 4 of
the present paper more supporting evidence is provided for this identification.

\subsection{Resolution of the ambiguity and explicit form factor solutions}

Based on the preceding considerations the ambiguity in the boundary
form factor solution can be fixed by requiring that\begin{equation}
\tilde{Q}_{n}=P_{n}^{(k)}\sigma_{n}\label{eq:maxdegree}\end{equation}
i.e.  \begin{eqnarray*}
Q_{n}(y_{1},\dots,y_{n}) & = & P_{n}^{(k)}(y_{1},\dots,y_{n})\sigma_{n}(y_{1},\dots,y_{n})\\
 &  & +\mbox{ terms of degree strictly less than }n(n+1)/2\end{eqnarray*}
For such a form factor the limiting procedure (\ref{eq:limitingff})
gives $\tilde{F}_{n}^{(k)}$. Therefore the limiting two-point function
cumulants defined in (\ref{eq:hbarlimit}) will be the same as for
the bulk operator $\mathrm{e}^{kg\Phi}$. By comparing the cumulant
expansion (\ref{eq:cumulantexpansion}) to its bulk counterpart, this
means that the boundary operator corresponding to this solution has
the conformal weight \[
h_{k}=-\frac{k^{2}g^{2}}{8\pi}\]
(at least to lowest order in $g^{2}$).

The boundary exponential field $\mathrm{e}^{\alpha\Phi(t,0)}$ has
conformal weight \[
-\frac{\alpha^{2}}{2\pi}\]
which leads to the conjecture that the boundary form factors satisfying
(\ref{eq:maxdegree}) correspond to the operators\[
\mathrm{e}^{\frac{kg}{2}\Phi}\]
The factor of one-half in the exponent is consistent with the form
of the boundary perturbation in the action (\ref{eq:boundaryaction}),
where a bulk perturbation which is a combination of $\mathrm{e}^{\pm g\Phi}$
is accompanied by a boundary perturbation which is a combination of
$\mathrm{e}^{\pm g\Phi/2}$.

The recurrence relations (\ref{eq:sinhgrecursions}) can be solved
as follows. The one-particle form factor is uniquely given by\[
Q_{1}^{(k)}(y_{1})=P_{1}^{(k)}\sigma_{1}(y_{1})=[k]y_{1}\]
The boundary kinematical recursion coefficient (\ref{eq:Bequ}) can
be expressed in terms of the elementary symmetric polynomials $\sigma_{l}$
in the form\[
B_{n}(y_{1},\dots,y_{n})=-\sum_{l=0}^{\left[\frac{n-1}{2}\right]}\left(2\cos\frac{\pi B}{2}\right)^{2l}\sigma_{n-1-2l}\]
Supposing that the solution up to $n-1$ particle level is known,
one can write the following Ansatz for the $n$-particle form factor\begin{equation}
Q_{n}^{(k)}=\epsilon_{1}\epsilon_{2}B_{n-1}Q_{n-1}^{(k)}+\sigma_{n}P_{n}^{(k)}+\sigma_{n}A_{n}^{(k)}\label{eq:ffansatz}\end{equation}
where $A_{n}^{(k)}$ is a general linear combination of products of
$\sigma_{l}$ with total degree strictly less than $n(n-1)/2$, and
the first term is understood with the replacement\[
\sigma_{l}^{(n-1)}\rightarrow\sigma_{l}^{(n)}\]
and the notation\[
\epsilon_{1}=2\cos\gamma\qquad,\qquad\epsilon_{2}=2\cos\gamma'\]
was introduced. The Ansatz (\ref{eq:ffansatz}) automatically solves
the boundary kinematical recursion $\mathcal{B}$ in (\ref{eq:sinhgrecursions})
and therefore the coefficients of $A_{n}^{(k)}$ can be determined
from the bulk kinematical recursion $\mathcal{K}$. Since the kernel
of the recursion is already fixed by (\ref{eq:ffansatz}), the solution
is unique. It is given by \begin{eqnarray*}
A_{2}^{(k)} & = & 0\\
A_{3}^{(k)} & = & [k](\epsilon_{1}^{2}+\epsilon_{2}^{2}-[k]\epsilon_{1}\epsilon_{2})\sigma_{1}\\
A_{4}^{(k)} & = & 4\sin^{2}\frac{\pi B}{2}\,[k]\sigma_{1}\sigma_{3}+[k]^{2}(\epsilon_{1}^{2}+\epsilon_{2}^{2}-[k]\epsilon_{1}\epsilon_{2})\sigma_{1}^{2}\left(\sigma_{2}+4\sin^{2}\frac{\pi B}{2}\right)\end{eqnarray*}
up to $4$-particle level, and it can easily be extended to higher
levels using any symbolic algebra software.

\section{Evaluating the cumulant expansion}

To provide evidence that the solution indeed corresponds to the exponential
operators as stated, let us study the short-distance behaviour of
the two-point correlators arising from it. It was already mentioned
that Koubek and Mussardo calculated the ultraviolet scaling dimension
to order $g^{2}$ by studying the short-distance limit of the two-point
function \cite{koubek_mussardo}; Babujian and Karowski evaluated
it from the cumulant expansion up to two-particle level \cite{babujiankarowski}.
Here the most singular term in the short-distance operator product
of any two of the operators in the family of minimal solutions is
expanded in powers of $g^{2}$ to order $g^{6}$. In addition, the
normalization of these operators is also evaluated to order $g^{2}$.

The conformal operator product algebra gives the leading term in the
short-distance limit of the two-point function as\[
\langle0|\mathrm{e}^{k\frac{g}{2}\Phi(\tau)}\mathrm{e}^{l\frac{g}{2}\Phi(0)}|0\rangle\sim\frac{1}{\tau^{2\Delta_{kl}}}\langle0|\mathrm{e}^{(k+l)\frac{g}{2}\Phi(0)}|0\rangle+\dots\]
with\begin{equation}
2\Delta_{kl}=-\frac{k^{2}g^{2}}{8\pi}-\frac{l^{2}g^{2}}{8\pi}+\frac{(k+l)^{2}g^{2}}{8\pi}=\frac{klg^{2}}{4\pi}\label{eq:cft_prediction}\end{equation}
where the standard conformal normalization for the exponential operators
was assumed. Recall that the operator thought to correspond to the
form factor solution in the previous section is normalized to have
unit vacuum expectation value, i.e. it takes the form \[
V_{k}(\tau)=\frac{1}{\langle0|\mathrm{e}^{k\frac{g}{2}\Phi(\tau)}|0\rangle}\mathrm{e}^{k\frac{g}{2}\Phi(\tau)}\]
so the short-distance expansion can be written as\[
\langle0|V_{k}V_{l}|0\rangle\sim\frac{C_{kl}^{k+l}}{(m\tau)^{2\Delta_{kl}}}\langle0|V_{k+l}|0\rangle+\dots\]
where the structure constant $C_{kl}^{k+l}$ is given by the ratio
of the appropriate vacuum expectation values. 

The cumulant expansion can be used to evaluate both $\Delta_{kl}$
and $C_{kl}^{k+l}$ \cite{babujiankarowski}. The first one can be
expressed in terms the asymptotic cumulants $\tilde{c}_{k}$, and
as a result the calculation is valid both for the boundary and the
bulk case, since the asymptotic cumulants $\tilde{c}_{k}$ are the
same. This is due to the way the ambiguity in the boundary solution
was fixed using the bulk solution (see eqn. (\ref{eq:maxdegree})). 

On the other hand, the evaluation of $C_{kl}^{k+l}$ involves the
full cumulants $c_{k}$. In the bulk theory these are identical to
$\tilde{c}_{k}$, but for the boundary theory they differ from them.
This is consistent with the fact that the $\tilde{c}_{k}$ are independent
of the boundary parameters, while the vacuum expectation values of
boundary fields depend on them in a very nontrivial way; therefore
the appearance the full cumulant $c_{k}$ is reassuring. Below it
is demonstrated that to order $g^{2}$ the result from the cumulant
expansion agrees with the exact vacuum expectation values conjectured
for the bulk theory in \cite{bulkvev}  and for the boundary case
in \cite{boundaryvev}.

\subsection{Evaluation of $\Delta_{kl}$}

Using the cumulant expansion (\ref{eq:cumulantexpansion}), $\Delta_{kl}$
can be evaluated in a systematic expansion in powers of $g^{2}$.
The $\tilde{c}_{n}$ can be obtained as cumulants of the asymptotic
spectral functions\[
\tilde{f}_{n}^{kl}\left(\theta_{1},\dots,\theta_{n}\right)=\lim_{\lambda\rightarrow\infty}f_{n}^{kl}\left(\theta_{1}+\lambda,\dots,\theta_{n}+\lambda\right)\]
which take the explicit form\[
\tilde{f}_{n}^{kl}\left(\theta_{1},\dots,\theta_{n}\right)=H_{n}^{2}P_{n}^{(k)}(x_{1},x_{2}\dots,x_{n})P_{n}^{(l)}(x_{1},x_{2}\dots,x_{n})\prod_{i<j}\frac{|f(\theta_{i}-\theta_{j})|^{2}}{(x_{i}+x_{j})^{2}}\quad,\quad x_{i}=\mathrm{e}^{\theta_{i}}\]
From (\ref{eq:fmin}) it is obvious that $f(\theta)=1+O(g^{2})$ for
any fixed $\theta\neq0$, and the leading term of the polynomials
$P_{n}^{(k)}$ is given by\begin{eqnarray*}
P_{n}^{(k)} & = & k\det m^{(n)}(k)+O(g^{2})\\
\mbox{where} &  & m_{ij}^{(n)}(k)=(i-j+k)\sigma_{2i-j}(x_{1},x_{2}\dots,x_{n})\quad i,j=1,\dots,n-1\end{eqnarray*}
while \[
H_{n}=\left(\frac{4\sin\pi B/2}{f(i\pi)}\right)^{n/2}\sim\left(\frac{g^{2}}{2}\right)^{n/2}+O(g^{n+1})\]
As a result \[
\tilde{f}_{n}^{kl}\sim g^{2n}\quad\Rightarrow\quad\tilde{c}_{n}^{kl}\sim g^{2n}\]
so an expansion up to order $g^{2n}$ only requires collecting the
appropriate terms from cumulants with up to $n$ particles. The first
cumulant can be easily expanded as\begin{equation}
\tilde{c}_{1}=H_{1}^{2}[k]^{2}\sim\!\frac{g^{2}}{2}kl+g^{6}\left(\frac{kl}{3072}-\frac{kl}{256\pi}-\frac{k^{3}l+kl^{3}}{768}\right)+O(g^{10})\label{eq:c1tog6}\end{equation}
and using (\ref{eq:1stcumulant}) gives \[
2\Delta_{kl}\sim\frac{g^{2}}{4\pi}kl+O(g^{6})\]
Therefore to recover the intended result it must be shown that all
higher order contributions cancel, which is demonstrated below up
to order $g^{6}$.

At order $g^{4}$ the second asymptotic cumulant is needed. It takes
the form\[
\tilde{c}_{2}(\theta_{1},\theta_{2})=\frac{1}{4}g^{4}k^{2}l^{2}\left(\left|f(\theta_{1}-\theta_{2})^{2}\right|-1\right)\]
At first this seems to be of order $O(g^{6})$, but some care must
be taken. Although it is true that $f(\theta)=1+O(g^{2})$ for any
fixed $\theta\neq0$, it also holds that $f(0)=0$. Therefore the
expansion of $f(\theta)$ in powers of $g^{2}$ does not converge
uniformly in the vicinity of the origin, and the series expansion
cannot be exchanged with the integration. However, the required integral
can be evaluated exactly \cite{babujiankarowski}; it has the following
small coupling expansion%
\footnote{In the formula derived in \cite{babujiankarowski} there seems to
be an overall sign mistake, which is corrected here.%
}\[
\int_{0}^{\infty}d\theta\left(\left|f(\theta)^{2}\right|-1\right)=-\frac{\pi^{2}}{24}B^{2}+O(B^{3})=O(g^{4})\]
and therefore the two-particle contribution is actually of order $O(g^{8})$.

The third cumulant can be written as\[
\tilde{c}_{3}(\theta_{1},\theta_{2},\theta_{3})=\frac{g^{6}}{8}(k^{3}l+kl^{3})\frac{\sigma_{3}}{\sigma_{3}-\sigma_{1}\sigma_{2}}+\frac{g^{6}}{8}kl\left(\frac{\sigma_{3}}{\sigma_{3}-\sigma_{1}\sigma_{2}}\right)^{2}+O(g^{8})\]
where the $\sigma_{k}$ are elementary symmetric polynomials of the
variables $x_{i}=\mathrm{e}^{\theta_{i}}$. The integral can be evaluated
as \[
\frac{1}{6}\int_{-\infty}^{\infty}\frac{d\theta_{1}}{2\pi}\int_{-\infty}^{\infty}\frac{d\theta_{2}}{2\pi}\tilde{c}_{3}(\theta_{1},\theta_{2},\theta_{3})=g^{6}\left(\frac{k^{3}l+kl^{3}}{768}+\frac{kl}{256\pi}-\frac{kl}{3072}\right)\]
which indeed cancels with the $g^{6}$ term in (\ref{eq:c1tog6}).

\subsection{Evaluation of $C_{kl}^{k+l}$}

$C_{kl}^{k+l}$ is the constant contribution to the logarithm of the
two-point function in the short-distance limit. It can be expressed
as\[
C_{kl}^{k+l}=\frac{\mathcal{G}(k+l,g)}{\mathcal{G}(k,g)\mathcal{G}(l,g)}\]
where $\mathcal{G}(k,g)$ denotes the exact vacuum expectation value
computed in \cite{bulkvev}:\begin{eqnarray*}
\mathcal{G}(k,g)=m^{2a^{2}}\left\langle \mathrm{e}^{a\varphi}\right\rangle  & = & \left[\frac{\Gamma\left[\frac{1}{2+2b^{2}}\right]\Gamma\left[1+\frac{b^{2}}{2+2b^{2}}\right]}{4\sqrt{\pi}}\right]^{-2a^{2}}\times\\
 &  & \exp\left\{ \int_{0}^{\infty}\frac{dt}{t}\left[-\frac{\sinh^{2}(2abt)}{2\sinh^{2}(bt)\sinh(t)\cosh((1+b^{2})t)}+2a^{2}\mathrm{e}^{-2t}\right]\right\} \end{eqnarray*}
and the parameters can be expressed in our notations as\[
b=\frac{g}{\sqrt{8\pi}}\quad,\quad a=\frac{kg}{\sqrt{8\pi}}\]
For small $g$ it can be written as\[
\mathcal{G}(k,g)\sim\exp\left(-\frac{k^{2}g^{2}}{4\pi}(-\log2+\gamma_{E})\right)\]
and so\begin{equation}
C_{kl}^{k+l}\sim\exp\left(-\frac{kl\, g^{2}}{2\pi}(-\log2+\gamma_{E})\right)\label{eq:bulkcpred}\end{equation}
On the other hand, it can easily be evaluated from the cumulant expansion
in the one-particle approximation. For the bulk theory, the first
term in (\ref{eq:1stcum}) vanishes exactly and so%
\footnote{A factor of two was inserted to compensate for the difference between
the bulk and boundary rapidity integrals.%
}\[
\log C_{kl}^{k+l}=2\frac{\tilde{c}_{1}}{2\pi}(\log2-\gamma_{E})+\dots=\frac{kl\, g^{2}}{2\pi}(\log2-\gamma_{E})+O(g^{4})\]
which agrees perfectly with (\ref{eq:bulkcpred}).

Turning now to the boundary case, the exact vacuum expectation values
of boundary fields in sine-Gordon theory were conjectured in \cite{boundaryvev}.
With the following action\begin{equation}
\mathcal{A}_{\mathrm{bsG}}=\int d^{2}x\left[\frac{1}{4\pi}(\partial_{t}\phi)^{2}-\frac{1}{4\pi}(\partial_{x}\phi)^{2}-2\mu\cos2\beta\phi\right]-2\mu_{B}\int dt\cos\beta(\phi(t,0)-\phi_{0})\label{eq:actionbsg}\end{equation}
the boundary one-point functions can be written in the following form:\begin{equation}
\left\langle \mathrm{e}^{ia\phi(t,0)}\right\rangle =\left(\frac{\pi\mu}{\Gamma(\beta^{2})}\right)^{\frac{a^{2}}{2(1-\beta^{2})}}g_{0}(a,\beta)g_{S}(a,\beta)g_{A}(a,\beta)\label{eq:exactbexpvev}\end{equation}
where%
\footnote{The integral in the exponent of $g_{S}$ contains an additional factor
of $1/2$ compared to the formula in \cite{boundaryvev}. The need
for such a factor was already noted in \cite{our_uvir}, and it was
also confirmed in a private discussion with Al.B. Zamolodchikov.%
}\begin{eqnarray*}
g_{0}(a,\beta) & = & \exp\left\{ \int_{0}^{\infty}\frac{dt}{t}\left[\frac{2\sinh^{2}(a\beta t)\left(\mathrm{e}^{(1-\beta^{2})t/2}\cosh(t/2)\cosh(\beta^{2}t/2)-1\right)}{\sinh(\beta^{2}t)\sinh(t)\sinh((1-\beta^{2})t)}-a^{2}\mathrm{e}^{-t}\right]\right\} \\
g_{S}(a,\beta) & = & \exp\left\{ \int_{0}^{\infty}\frac{dt}{t}\frac{\sinh^{2}(a\beta t)\left(2-\cos(2zt)-\cos(2\bar{z}t)\right)}{2\sinh(\beta^{2}t)\sinh(t)\sinh((1-\beta^{2})t)}\right\} \\
g_{A}(a,\beta) & = & \exp\left\{ \int_{0}^{\infty}\frac{dt}{t}\frac{\sinh(2a\beta t)\left(\cos(2zt)-\cos(2\bar{z}t)\right)}{\sinh(\beta^{2}t)\sinh(t)\cosh((1-\beta^{2})t)}\right\} \end{eqnarray*}
and%
\footnote{In sine-Gordon theory, $\bar{z}$ is the complex conjugate of $z$,
but this does not remain true under analytic continuation to sinh-Gordon
theory.%
}\begin{equation}
\cosh^{2}\pi z=\mathrm{e}^{-2i\beta\phi_{0}}\frac{\mu_{B}^{2}\sin\pi\beta^{2}}{\mu}\qquad,\qquad\cosh^{2}\pi\bar{z}=\mathrm{e}^{2i\beta\phi_{0}}\frac{\mu_{B}^{2}\sin\pi\beta^{2}}{\mu}\label{eq:zuvir}\end{equation}
provided the operators are normalized as\begin{eqnarray*}
\mathrm{e}^{2ia\phi(x)}\mathrm{e}^{-2ia\phi(x)} & \sim & \frac{1}{\left|x-y\right|^{4a^{2}}}+\dots\\
\mathrm{e}^{ia\phi(t,0)}\mathrm{e}^{-ia\phi(t',0)} & \sim & \frac{1}{\left|t-t'\right|^{2a^{2}}}+\dots\end{eqnarray*}
The coupling $\mu$ can be related to the mass $m$ of the lightest
breather as follows \cite{mass_scale}:\[
\mu=\frac{\Gamma(\beta^{2})}{\pi\Gamma(1-\beta^{2})}\left[\frac{m}{2\sin\frac{\pi\beta^{2}}{2(1-\beta^{2})}}\frac{\sqrt{\pi}\Gamma\left(\frac{1}{2(1-\beta^{2})}\right)}{2\Gamma\left(\frac{\beta^{2}}{2(1-\beta^{2})}\right)}\right]^{2-2\beta^{2}}\]
Eqn. (\ref{eq:exactbexpvev}) can be expanded for small $\beta$ as
\[
\mathcal{G}_{B}(\kappa,\beta)=m^{-2\kappa^{2}\beta^{2}}\left\langle \mathrm{e}^{i\kappa\beta\phi(t,0)}\right\rangle \sim\mathcal{N}(\kappa,\beta)\bar{g}_{0}(\kappa,\beta)\bar{g}_{S}(\kappa,\beta)\bar{g}_{A}(\kappa,\beta)\]
with\begin{eqnarray*}
\mathcal{N}(\kappa,\beta) & = & 1-2\kappa^{2}\beta^{2}\log2+O(\beta^{4})\\
\bar{g}_{0}(\kappa,\beta) & = & 1+\kappa^{2}\beta^{2}(1+\gamma_{E}+\log2)+O(\beta^{4})\\
\bar{g}_{S}(\kappa,\beta) & = & 1+\kappa^{2}\beta^{2}\left(\frac{1}{2}\pi z\coth\pi z+\frac{1}{2}\pi\bar{z}\coth\pi\bar{z}-1\right)+O(\beta^{4})\\
\bar{g}_{A}(\kappa,\beta) & = & 1+\kappa\int_{0}^{\infty}\frac{dt}{t}\left(\cos(2zt)-\cos(2\bar{z}t)\right)\left(\frac{4}{\sinh[2t]}+\frac{2t}{\cosh[t]^{2}}\beta^{2}\right)+O(\beta^{4})\end{eqnarray*}
Due to linearity of $\bar{g}_{A}$ in $\kappa$ its contribution drops
from $C_{kl}^{k+l}$, so it is not necessary to evaluate it explicitly.
The sinh-Gordon expectation values can be obtained by taking the analytic
continuation $\beta\rightarrow ig/\sqrt{8\pi}$, under which the operator\[
\mathrm{e}^{k\frac{g}{2}\Phi(\tau)}\]
corresponds to putting $\kappa=k$. Collecting all the terms the final
expression is \begin{eqnarray}
\log C_{kl}^{k+l} & = & \log\left.\frac{\mathcal{G}_{B}(k+l,\beta)}{\mathcal{G}_{B}(k,\beta)\mathcal{G}_{B}(l,\beta)}\right|_{\beta\rightarrow ig/\sqrt{8\pi}}\nonumber \\
 & = & -\frac{klg^{2}}{4\pi}(-\log2+\gamma_{E}+\frac{1}{2}\pi z\coth\pi z+\frac{1}{2}\pi\bar{z}\coth\pi\bar{z})+O(g^{4})\label{eq:blogcfromexact}\end{eqnarray}
On the other hand, from the cumulant expansion\[
\log C_{kl}^{k+l}=\frac{\tilde{c}_{1}}{2\pi}(\log2-\gamma_{E})+\int_{0}^{\infty}\frac{d\theta}{2\pi}\left(c_{1}(\theta)-\tilde{c}_{1}\right)\]
From (\ref{eq:rmin}) it is easy to obtain that\[
r(\theta)=\frac{i\sinh\theta(1-i\sinh\theta)}{(\sinh\theta-i\sin\gamma)(\sinh\theta-i\sin\gamma')}+O(g^{2})\]
and so \[
\int_{0}^{\infty}\frac{d\theta}{2\pi}\left(c_{1}(\theta)-\tilde{c}_{1}\right)=\frac{klg^{2}}{4\pi}\int_{0}^{\infty}d\theta\left\{ \frac{\sinh^{2}\theta(1+\sinh^{2}\theta)}{(\sinh^{2}\theta+\sin^{2}\gamma)(\sinh^{2}\theta+\sin^{2}\gamma')}-1\right\} \]
The integral can be evaluated using \[
\int_{0}^{\infty}\frac{d\theta}{\sinh^{2}\theta+\sin^{2}\gamma}=\frac{\pi-2|\gamma|}{|\sin2\gamma|}\quad\mbox{for}\quad-\frac{\pi}{2}<\gamma<\frac{\pi}{2}\]
For definiteness let us suppose that \[
-\frac{\pi}{2}<\gamma,\gamma'<0\]
which means that $0<E,F<1$ as a consequence of the parameter identification
in (\ref{eq:rmin}). This leads to\begin{eqnarray}
\log C_{kl}^{k+l} & = & \frac{klg^{2}}{4\pi}\big[\log2-\gamma_{E}-\frac{1}{2}(\pi+\gamma+\gamma')\cot(\pi+\gamma+\gamma')\label{eq:blogcfromcumul}\\
 &  & -\frac{1}{2}(\gamma-\gamma')\cot(\gamma-\gamma')\big]\nonumber \\
 & = & \frac{klg^{2}}{4\pi}\big[\log2-\gamma_{E}-\frac{1}{4}\pi(E+F)\cot\frac{1}{2}\pi(E+F)\nonumber \\
 &  & -\frac{1}{4}\pi(E-F)\cot\frac{1}{2}\pi(E-F)\big]\nonumber \end{eqnarray}
The expressions (\ref{eq:blogcfromexact}) and (\ref{eq:blogcfromcumul})
match provided\begin{equation}
z=\frac{1}{2}(E-F)\quad,\quad\bar{z}=\frac{1}{2}(E+F)\label{eq:zEFmatch}\end{equation}
Some care must be taken because there is a branch choice ambiguity
in the $z$, $\bar{z}$ parametrization which is obvious from (\ref{eq:zuvir});
in (\ref{eq:zEFmatch}) an implicit choice was made. It is only necessary
to show that there exists one particular choice which is consistent;
the correspondence on the other branches (i.e. outside the region
$0<E,F<1$) can be obtained by analytic continuation. 

The consistency condition is that (\ref{eq:zEFmatch}) must provide
an identification of the reflection factor \begin{equation}
R(\theta)=\left(\frac{1}{2}\right)_{\theta}\left(\frac{1}{2}+\frac{B}{4}\right)_{\theta}\left(1-\frac{B}{4}\right)_{\theta}\left[\frac{E-1}{2}\right]_{\theta}\left[\frac{F-1}{2}\right]_{\theta}\label{eq:bshgrefl_copy}\end{equation}
(cf. (\ref{eq:shgrefl})) of the sinh-Gordon particle to the that
of the first sine-Gordon breather under the analytic continuation
$\beta\rightarrow ig/\sqrt{8\pi}$. The reflection factor of the first
breather can be written as \cite{ghoshal}\begin{eqnarray}
R^{(1)}(\vartheta) & = & \left(\frac{1}{2}\right)_{\theta}\left(\frac{1}{2}-\frac{\xi}{2}\right)_{\theta}\left(1+\frac{\xi}{2}\right)_{\theta}\left[\frac{\xi\eta}{\pi}-\frac{1}{2}\right]_{\theta}\left[\frac{i\xi\vartheta}{\pi}-\frac{1}{2}\right]_{\theta}\label{b1_refl}\\
 &  & \xi=\frac{\beta^{2}}{1-\beta^{2}}\nonumber \end{eqnarray}
which can be compared with (\ref{eq:shgrefl}). Under the analytic
continuation $\beta\rightarrow ig/\sqrt{8\pi}$ \[
B=-2\xi\]
and so the factors independent of the boundary condition agree. From
the following relation between the Lagrangian and bootstrap parameters\begin{eqnarray*}
\cos\left(\beta^{2}\eta\right)\cosh\left(\beta^{2}\vartheta\right) & = & \mu_{B}\sqrt{\frac{\sin\pi\beta^{2}}{\mu}}\cos\left(\beta\phi_{0}\right)\\
\sin\left(\beta^{2}\eta\right)\sinh\left(\beta^{2}\vartheta\right) & = & \mu_{B}\sqrt{\frac{\sin\pi\beta^{2}}{\mu}}\sin\left(\beta\phi_{0}\right)\end{eqnarray*}
derived by Al.B. Zamolodchikov \cite{sinG_uvir} (see also \cite{our_uvir}),
comparing to (\ref{eq:zuvir}) the following relation is obtained:\begin{equation}
z=\frac{\beta^{2}}{\pi}(\eta-i\vartheta)\quad,\quad\bar{z}=\frac{\beta^{2}}{\pi}(\eta+i\vartheta)\label{eq:zetathetamatch}\end{equation}
where an appropriate choice of branches is understood. To the leading
order%
\footnote{Since eqn. (\ref{eq:zEFmatch}) itself is only valid to leading order
in $\beta^{2}$, it doesn't make sense to take the comparison beyond
this approximation.%
} in $\beta^{2}$ (\ref{eq:zEFmatch}) and (\ref{eq:zetathetamatch})
are equivalent to \[
E=\frac{2\xi\eta}{\pi}\quad,\quad F=\frac{2i\xi\vartheta}{\pi}\]
which makes the reflection factors (\ref{eq:bshgrefl_copy}) and (\ref{b1_refl})
identical.

\section{Discussion and outlook}

With the method developed in this paper, it proved possible to organize
the infinitely many minimal boundary form factor solutions of sinh-Gordon
theory encountered in \cite{bffcount} into well-defined families
by restricting the highest-growing term to be equal to the bulk form
factor of a given exponential operator. It was shown that such a solution
has the correct scaling dimension to order $g^{6}$, and the claim
that its vacuum expectation value is normalized to $1$ matches with
the exact vacuum expectation values conjectured in \cite{boundaryvev}
to order $g^{2}$.

The restriction leading to unique determination of the solution was
made possible by the observation made in \cite{bffcount} that the
dominant asymptotic contribution of the boundary form factor always
solves the bulk form factor equations. It is very important to note
that the construction always guarantees that the cumulant expansion
for the scaling dimension of the boundary operator coincides with
that of the bulk one which was used as an input for the solution of
the boundary form factor bootstrap axioms. 

This observation leads to a proposal for constructing form factors
of relevant boundary operators. In the bulk, form factors of spinless
relevant fields can be selected from the space of solutions to the
form factor axioms by the asymptotic factorization condition (\ref{eq:ffcluster})
which was derived in \cite{delta-th}. Using these solutions as an
input one can then construct solutions for relevant boundary fields
whose scaling dimension coincides with the right (or equally, the
left) conformal dimension of the corresponding bulk field. In fact,
the form factors of the only primary boundary field in the Lee-Yang
model computed in \cite{bffprogram,bfftcsa} could also have been
derived this way.

In principle, the agreement of the most singular term calculated from
the form factor expansion with the quantum field expectation for some
operator $\mathcal{O}$ leaves open the possibility of mixing with
other fields of smaller weight (i.e. which are more relevant) in the
sense that it allows that the form factor corresponds to $\mathcal{O}$
with an addition of an unspecified linear combination of operators
with smaller weight, i.e. to an operator of the form \[
\mathcal{O}+\sum_{l}C_{l}\mathcal{O}_{l}\]
where the $\mathcal{O}_{l}$ are operators with scaling dimension
less than that of $\mathcal{O}$. This problem was already noted in
\cite{bffcount}. However, for the most relevant boundary field this
issue is definitely not present which is reassuring for the determination
of its form factors using the construction proposed above.

In the sinh-Gordon case the spectrum of conformal dimensions is unbounded
from below, but the fact that the short-distance limit of the two-point
function constructed from form factors is consistent with the leading
part expected from the conformal operator product algebra and the
exact vacuum expectation values eliminates some possibilities of such
mixing. However, it still allows e.g. for\begin{equation}
\mathrm{e}^{k\frac{g}{2}\Phi(t,0)}+\sum_{l>k}C_{l}\mathrm{e}^{l\frac{g}{2}\Phi(t,0)}\label{eq:mixingexample}\end{equation}
(where it was assumed that $k>0$). 

Another restriction comes from the continuation to sine-Gordon theory.
The parameter $k$ of the solution which gives the corresponding boundary
exponential operator as \[
\mathrm{e}^{k\frac{g}{2}\Phi(t,0)}\]
can take any real value, but the operators with $k\in\mathbb{Z}$
form a closed operator algebra, which is the minimal one consistent
with the presence of the boundary potential in (\ref{eq:boundaryaction}).
Continuing analytically to the sine-Gordon theory with action (\ref{eq:actionbsg}),
the $n$-particle form factors of these operators become the form
factors of \[
\mathrm{e}^{ik\beta\phi(t,0)}\]
with a many-particle state that consists exclusively of $n$ first
breathers. The operators with $k=\pm1$ are the most relevant ones
in this family, and therefore they cannot mix with other operators
at all; this property is expected to hold after continuing back to
sinh-Gordon theory as well. In fact, these would be the operators
for which there is the widest spectrum of subleading operators to
choose from in (\ref{eq:mixingexample}), therefore the absence of
mixing in their case is particularly striking.

The above arguments cannot be considered complete, but taken together
they make it likely that there is eventually no operator mixing for
the exponential fields in the boundary sinh-Gordon model at all. The
basis provided by the solutions constructed in this paper is the only
one for which the following property holds: for any member of the
basis, the limiting form factor (in the sense of (\ref{eq:limitingff}))
is not only a solution of the bulk form factor axioms, but also satisfies
the asymptotic factorization condition (\ref{eq:ffcluster}). Note
that this factorization property is needed for the derivation of the
cumulant expansion (\ref{eq:cumulantexpansion}) and it also ensures
that the integrals in the expansion are convergent (cf. eqns. (\ref{eq:cumulantdecay})
and (\ref{eq:cumulantapproach})). Certainly some deeper understanding
of the boundary form factor bootstrap is needed in order to relate
this condition to properties of relevant boundary primary fields,
and to make the correspondence between form factor solutions and operators
more precise.

\subsection*{Acknowledgments}

I would like to thank F. Smirnov for a useful discussion, in particular
for drawing my attention to the cumulant expansion; also Z. Bajnok
and L. Palla for comments about the manuscript. This research was
partially supported by the Hungarian research fund OTKA K60040 and
also by a Bolyai J\'anos research scholarship.

\end{document}